\documentclass[journal]{IEEEtran}

\ifCLASSINFOpdf
\else
   \usepackage[dvips]{graphicx}
\fi
\usepackage{url}

\usepackage{graphicx}

\usepackage{bm}

\usepackage{cite}
\usepackage{amsmath,amssymb,amsfonts}
\usepackage{textcomp}

\begin{document}

\title{Deep Learning-Assisted Jamming Mitigation with Movable Antenna Array}

\author{Xiao Tang, Yudan Jiang, Jinxin Liu, Qinghe Du, Dusit Niyato, and Zhu Han
\thanks{X. Tang is with the School of Information and Communication Engineering, Xi'an Jiaotong University, Xi'an 710049, China. (e-mail: tangxiao@xjtu.edu.cn)}
\thanks{Y. Jiang is with the School of Electronics and Information, Northwestern Polytechnical University, Xi'an 710072, China.}
\thanks{J. Liu is with the School of Mechanical Engineering, Xi’an Jiaotong University, Xi’an 710049, China.}
\thanks{Q. Du is with the School of Information and Communication Engineering, Xi'an Jiaotong University, Xi'an 710049, China.}
\thanks{D. Niyato with the College of Computing and Data Science, Nanyang Technological University, Singapore.}
\thanks{Z. Han is with the Department of Electrical and Computer Engineering, University of Houston, Houston 77004, USA.}
}


\maketitle

\begin{abstract}
This paper reveals the potential of movable antennas in enhancing anti-jamming communication. We consider a legitimate communication link in the presence of multiple jammers and propose deploying a movable antenna array at the receiver to combat jamming attacks. We formulate the problem as a signal-to-interference-plus-noise ratio maximization, by jointly optimizing the receive beamforming and antenna element positioning. Due to the non-convexity and multi-fold difficulties from an optimization perspective, we develop a deep learning-based framework where beamforming is tackled as a Rayleigh quotient problem, while antenna positioning is addressed through multi-layer perceptron training. The neural network parameters are optimized using stochastic gradient descent to achieve effective jamming mitigation strategy, featuring offline training with marginal complexity for online inference. Numerical results demonstrate that the proposed approach achieves near-optimal anti-jamming performance thereby significantly improving the efficiency in strategy determination.
\end{abstract}


\IEEEpeerreviewmaketitle

\section{Introduction}

The widely existed malicious jamming has presented a major challenge for wireless information security, for which effective anti-jamming techniques have long been a critical requisite for wireless systems~\cite{jam,jamming}. The existing arts to combat jamming attacks often rely on jamming mitigation or avoidance by jointly exploiting multi-dimensional resources. However, as the wireless environments become more sophisticated and adversaries become more intricate, the implementation of these methods requires more intensive resources or computations~\cite{jam-1,jam-2}. This has driven research efforts seeking for new dimensions of resources or degrees of freedom to mitigate the jamming attacks for information security provisioning.

Recently, movable antennas have emerged as a promising technology with the potential to enhance wireless communications in various aspects~\cite{ma-1,ma-2}. Unlike conventional fixed antenna systems, movable antennas can dynamically adjust their spatial positions to actively control the radiation pattern in desired directions~\cite{ma-base}. This capability resembles reconfigurable intelligent surfaces that intentionally intervene the propagation, yet movable antennas can be more advantageous in terms of transmitter/receiver-based implementation without requiring additional deployment of surfaces~\cite{ris-1,ris-2}. Therefore, movable antenna technologies have attracted the research efforts in diverse perspectives, such as transmission improvement~\cite{tran-1,tran-2,tran-3,tran-4}, sensing enhancement~\cite{sen-1,sen-2}, computation assistance~\cite{comp-1,comp-2}, and security provisioning~\cite{sec-1,sec-2,sec-3}. In these studies, the flexible configurations of movable antennas are jointly exploited with resources in conventional dimensions to achieve superior system performance, which demonstrates the significant advantages to exploit the antenna movement as a new degree of freedom.

Despite the wide attention and emerging research efforts on movable antenna-assisted communications, its potential in jamming mitigation remains unrevealed. Inspired by existing results, the spatial adaptability of movable antennas empowers reliable communication even in highly dynamic and hostile environments, thereby achieving effective jamming mitigation beyond the capability of conventional fixed antenna systems. Furthermore, by jointly exploiting the antenna movement with other resources, the improved spatial diversity allows to find appropriate paths to circumvent malicious attacks, and thus alleviates jamming even when involving multiple adversaries. 

Moreover, the existing work mostly adopts optimization-based analysis for movable antenna-based communications. However, the antenna positions are incorporated in the radiation, and consequently the treatment can be rather involved. As such, recent studies also refer to learning techniques for more efficient solutions~\cite{nn,nn-1,cja}. In this regard, machine learning as a powerful tool for tackling complex, non-convex optimization, can efficiently determine the (near-)optimal configuration for both antenna positions and transmission strategy, providing real-time adaptability that traditional optimization methods struggle to achieve. Therefore, by integrating movable antenna transmissions with learning-based scheme design, a highly adaptive and resilient anti-jamming system can be anticipated to effectively fit the dynamic wireless environments.

Consequently, in this paper, we explore the potential of the movable antennas in anti-jamming communications, and propose a deep learning-based framework that learns the optimal antenna positions and beamforming coefficients, achieving superior jamming mitigation. Particularly, we investigate an adversarial scenario with multiple jammers, and the countermeasure with movable antenna-based reception is proposed to mitigate jamming attacks and maximize the transmission signal-to-interference-plus-noise ratio (SINR). To address the non-convexity of the problem, we propose the decomposition such that receive beamforming is obtained through Rayleigh quotient and antenna positioning is approximated with a multi-layer perceptron (MLP). Within a learning framework, the two subproblems are interacted and jointly trained for effective anti-jamming strategy output, which facilitates efficient online inference. Simulation results validate the proposed approach, demonstrating its near-optimal performance in jamming mitigation and the advantages in terms of computation efficiency.

\section{System Model}

\begin{figure}[t]
   \centering
   \includegraphics[width=0.5\textwidth]{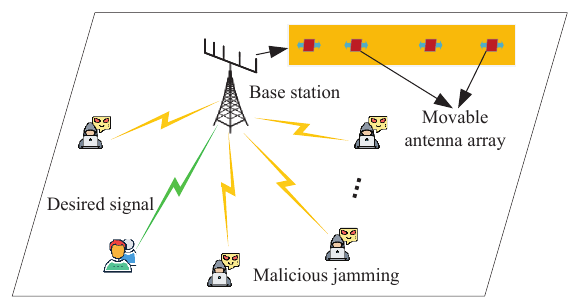} 
   \caption{System model.}
   \label{fig:sys}
\end{figure}

We consider a wireless system with a legitimate communication pair in the presence of jamming attacks, as illustrated in Fig.~\ref{fig:sys}. The single-antenna legitimate source node transmits a desired signal to a multi-antenna base station. There are $K$ signal-antenna adversaries, the set of which is denoted by $\mathcal{K} = \left\{ 1, 2, \cdots, K \right\}$, emitting jamming signals over the same frequency band to deteriorate the legitimate transmissions. To defend against the jamming attacks, the antenna at the base station is enabled with flexible movement as a linear movable antenna array to facilitate effective jamming mitigation. Particularly, the movable antenna array consists of $ N $ elements, denoted as $\mathcal{N} = \left\{ 1, 2, \cdots, N \right\}$, whose positions are denoted as $ \bm{x} = \left[x_n\right]_{n\in\mathcal{N}} $ within the range $ \left[0, L\right] $ as $ L $ denoting the size of allowed region.

Based on the locations of the legitimate source and jamming nodes in the considered system, the directions of arrival of the their transmitted signals at the base station are denoted as $ \theta_0 $ for the legitimate source and $ \left\{ \theta_k \right\}_{k\in\mathcal{K}} $ for the $ K $ jammers. In this work, we particularly address the movable antenna-enabled jamming resolution in terms of directions and adopt the field-response based channel model, similar as~\cite{tran-3,sec-1,nn-1}. Accordingly, the steering vectors from the transmitters to the antenna array at the base station are specified as
\begin{equation}
   \bm{a}\left(\bm{x}, \theta_k\right) = \left[ e^{j\frac{2\pi}{\lambda}x_n\cos\theta_k} \right]_{n\in\mathcal{N}}, \quad \forall k\in\{0\}\cup\mathcal{K},
\end{equation}
where $ \lambda $ denotes the wavelength of the carrier, and the index-0 refers to the legitimate user. Meanwhile, the movable antenna-based base station conducts receive beamforming over the aggregated receive signal, denoted as $ \bm{w} $. Then, the SINR of the received signal is obtained as
\begin{equation}
   \eta = \frac{ \left| \bm{w}^H \bm{a}\left(\bm{x}, \theta_0\right) \right|^2 } { \sum\limits_{k\in\mathcal{K}} \left| \bm{w}^H \bm{a}\left(\bm{x}, \theta_k\right) \right|^2 + \sigma_0^2  },
\end{equation}
where $ \sigma_0^2 $ denotes the background noise power.

For the considered system, we intend to defend against the malicious jamming attacks so as to protect the desired legitimate transmissions. Accordingly, we resort to jointly investigate the receive beamforming as well as the antenna element positioning such that the legitimate reception can be well focused while the adversary jamming can be sufficiently suppressed. In this regard, the flexible movement of the antenna allows us to exploit beyond the conventional spatial diversity with elaborate radiation pattern to achieve this goal. Therefore, the problem is formulated to maximize the receive SINR with receive beamforming and antenna position optimization as
\begin{IEEEeqnarray}{cl}\label{eq:opt}
   \IEEEyesnumber \IEEEyessubnumber*
   \max_{ \bm{x},\bm{w} } &\quad \eta \\
   \mathrm{s.t.}
   &\quad x_n \in \left[0, L \right],\quad \forall n\in\mathcal{N}, \label{eq:x-con-1}\\
  &\quad \left| x_n - x_{n-1} \right| \ge d_{\min}, \quad \forall n\in\mathcal{N} \backslash \{1\}, \label{eq:x-con-2} \\
   &\quad \left\| \bm{w} \right\| = 1, \label{eq:w-con}
\end{IEEEeqnarray}
where $ d_{\min} $ is the minimum antenna spacing to prevent coupling effect between adjacent antenna elements and satisfies $ \left(N-1\right)d_{\min} \le L $ without loss of generality, and $ \bm{w} $ as the receive beamforming is naturally normalized. While this study considers a single-user scenario to simplify the analysis, the proposed framework can be conveniently extended to multi-user environments. In such cases, we can endeavor for the maximization of the minimum SINR overall all users or extend to the transmission rate to maximize the system throughput accordingly.

Generally, the problem in~(\ref{eq:opt}) is non-convex and thus difficult to solve effectively. Specifically, besides the non-convex objective, the constraints in~(\ref{eq:x-con-2}) and~(\ref{eq:w-con}) induce some non-convex regions that are intricate to deal with. Also, the antenna element positioning is involved within complex exponential functions which are difficult to tackle. Moreover, the beamforming and antenna positioning are mutually influenced and the combined effect needs to be analyzed so as to achieve the optimized performance.

\begin{figure*}[!t] \vspace{5pt}
   \centering
   \includegraphics[width=0.75\textwidth]{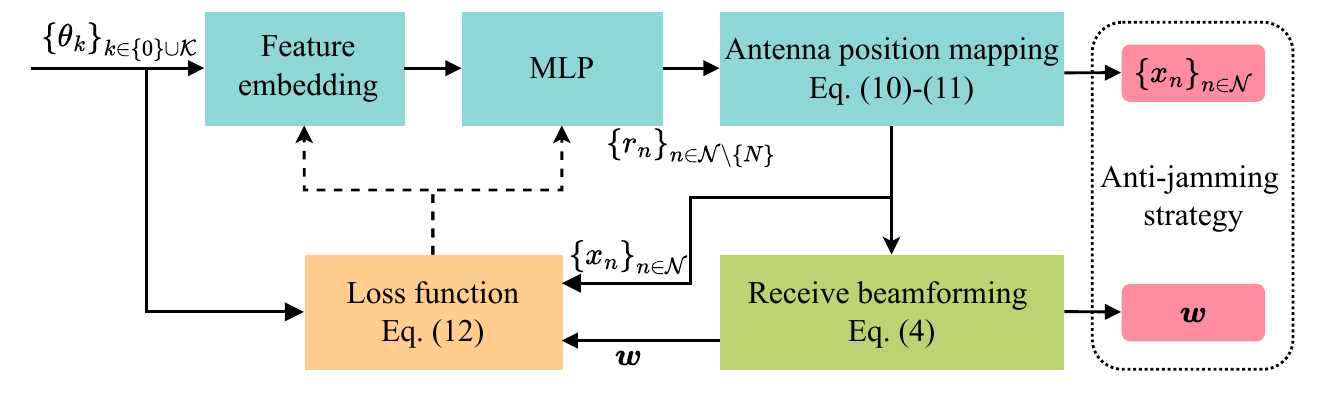}
   \caption{The proposed learning architecture.}
   \label{fig:arch} 
\end{figure*}

\section{Learning-Assisted Approach}

Due to the non-convex nature of the formulated problem, the problem solving in conventional manners usually requires certain iterative approximation processes that can be rather time-consuming. Towards this issue, we resort to the deep learning approach for efficient algorithm design. The formulated problem involves optimizing antenna positioning and beamforming for a given set of directions of arrival, where the deterministic input-output relationships can be efficiently incorporated in a deep learning framework. Unlike deep reinforcement learning, which is more suited for sequential decision-making processes, deep learning provides a direct mapping from network input to the optimal configuration, ensuring computational efficiency and interpretability. Consequently, we propose the deep learning-assisted approach as follows.

\subsection{Learning Architecture}

Revisiting the jamming mitigation problem formulation in~(\ref{eq:opt}), we seek for a policy combination in terms of beamforming and antenna positioning with respect to a given directions of arrival of signals from the transmitters to achieve the highest SINR. Therefore, the overall architecture for problem solving takes the directions of arrival of signals as input and produces the anti-jamming policy. Furthermore, as the policy incorporates two parts, i.e., the receive beamforming and the antenna element positioning, we then analyze the properties of the corresponding subproblems and design the optimizers for them accordingly. In this respect, the computation efficiency and solution quality can be well balanced through such a decomposition with tailored optimizers.

In accordance with the individual optimizer design, the problem in~(\ref{eq:opt}) is decomposed into two subproblems as
\begin{IEEEeqnarray}{cl}\label{eq:opt-w}
   \IEEEyesnumber \IEEEyessubnumber*
   \max_{ \bm{w} } &\quad \eta = \frac{\bm{w}^H \bm{A}\bm{w}}{\bm{w}^H \bm{B}\bm{w}} \\
   \mathrm{s.t.} & \quad \text{(\ref{eq:w-con})},
\end{IEEEeqnarray}
and
\begin{IEEEeqnarray}{cl}\label{eq:opt-x}
   \IEEEyesnumber \IEEEyessubnumber*
   \max_{ \bm{x} } &\quad \eta \\
   \mathrm{s.t.} & \quad \text{(\ref{eq:x-con-1})}, \text{(\ref{eq:x-con-2})},
\end{IEEEeqnarray}
for the receive beamforming and antenna element positioning, respectively, where
\begin{IEEEeqnarray}{l}\label{eq:AB}
   \IEEEyesnumber \IEEEyessubnumber*
   \bm{A} = \bm{a}\left(\bm{x}, \theta_0\right) \bm{a}^H\left(\bm{x}, \theta_0\right), \\
   \bm{B} = \sum_{k\in\mathcal{K}} \bm{a}\left(\bm{x}, \theta_k\right) \bm{a}^H\left(\bm{x}, \theta_k\right) + \sigma_0^2 \bm{I}_N,
\end{IEEEeqnarray}
are introduced to facilitate the investigation of receive beamforming with $ \bm{I}_N $ being an $N$-dimensional identity matrix. Then, for the receive beamforming subproblem in~(\ref{eq:opt-w}), with $ \bm{A} $ and $ \bm{B} $ evidently being Hermitian matrices, it appears in the form of generalized Rayleigh quotient~\cite{rayq}. As such, to find the optimal beamforming for the highest SINR, we can conduct eigenvalue decomposition of $ \bm{B}^{-1}\bm{A} $ and adopt the normalized eigenvector corresponding to the largest eigenvalue as the desired receive beamformer. Here, since $ \bm{A} $ is obtained as the outer product of vector $ \bm{a}\left(\bm{x}, \theta_0\right) $ with its complex conjugate, we can follow the similar approach in~\cite[Proposition~3]{rayq}, and derive the optimal beamformer as
\begin{equation}
   \bm{w}^{\star} = \frac{\bm{B}^{-1}\bm{a}\left(\bm{x}, \theta_0\right)}{\| \bm{B}^{-1}\bm{a}\left(\bm{x}, \theta_0\right) \|},
\end{equation}
along with the largest eigenvalue, or the highest SINR as
\begin{equation}
   \eta(\bm{w}^{\star}) =  \bm{a}^H\left(\bm{x}, \theta_0\right)\bm{B}^{-1}\bm{a}\left(\bm{x}, \theta_0\right).
\end{equation}


Furthermore, for the antenna element positioning subproblem in~(\ref{eq:opt-x}), it is relatively complicated because of the complex exponential functions as well as the non-convex objective and feasible region. To address the difficulties from an optimization perspective, we adopt a deep neural network to learn the complicated relationship between the known directions of arrival and the desired antenna element positions. In this regard, the universal representation capability of neural networks can approximate the objective function to approach the highest SINR. A major challenge here is the non-convex feasible region in~(\ref{eq:opt-x}) due to the minimum required antenna spacing. To this issue, we introduce a series of intermediate variables $ \left\{ r_n \right\}_{n\in\mathcal{N}\backslash\{N\}} $ as the ratio between smallest spacing and the actually achieved spacing between adjacent antenna elements. Naturally, we then have $ r_n \in (0, 1] $, $ \forall n \in\mathcal{N}\backslash\{N\} $ that fits the general neural network output. Then, the tentative antenna element positions are obtained as
\begin{equation} \label{eq:tilde-x} \left\{
\begin{aligned}
   &\tilde{x}_1 = 0, \\
   &\tilde{x}_n = \tilde{x}_{n-1} + \frac{d_{\min}}{r_{n-1}} = d_{\min}\sum\limits_{m=1}^{n-1}\frac{1}{r_m}, \quad \forall n \in\mathcal{N}\backslash\{1\},
\end{aligned} \right.
\end{equation}
where the first antenna element is set at the initial position without loss of generality. We can see that the positioning in~(\ref{eq:tilde-x}) evidently satisfies the constraint in~(\ref{eq:x-con-1}). If the maximum allowed antenna range presented in~(\ref{eq:x-con-2}) is also satisfied, the results in~(\ref{eq:tilde-x}) can be used as the actual antenna positioning. Otherwise, the following treatment is conducted to ensure the antenna range constraint. For such cases, we rewrite the antenna spacing between adjacent elements as
\begin{equation}
   x_n - x_{n-1} = d_{\min} + \delta\left( \frac{1}{r_{n-1}} - 1 \right), \quad \forall n \in\mathcal{N}\backslash\{1\},
\end{equation}
where $ \delta\left( \frac{1}{r_{n-1}} - 1 \right) $ is the additional distance besides the minimum spacing with $ \delta = d_{\min} $ for the tentative results in~(\ref{eq:tilde-x}). Then, by letting $ x_N = L $ such that the last element is rearranged at the maximum allowed range, we derive that
\begin{equation}
   \delta = \frac{L - \left(N-1\right)d_{\min}}{\sum\limits_{n \in\mathcal{N}\backslash\{N\}}\left( \frac{1}{r_{n}} - 1 \right)},
\end{equation}
which thus guarantees the constraint in~(\ref{eq:x-con-2}). Therefore, based on the neural network output $ \left\{ r_n \right\}_{n\in\mathcal{N}\backslash\{N\}} $, the antenna element positions are constructed as
\begin{equation} \label{eq:x} \left\{
\begin{aligned}
   &x_1 = 0, \\
   &x_n = x_{n-1} + d_{\min} + \delta\left( \frac{1}{r_{n-1}} - 1 \right), \quad \forall n \in\mathcal{N}\backslash\{1\},
\end{aligned} \right.
\end{equation}
with
\begin{equation} \label{eq:delta} \delta = \left\{
\begin{aligned}
   & d_{\min}, &&\quad \text{if } x_N \le L \\
   & \frac{L - \left(N-1\right)d_{\min}}{\sum\limits_{n \in\mathcal{N}\backslash\{N\}}\left( \frac{1}{r_{n}} - 1 \right)}, &&\quad \text{otherwise},
\end{aligned} \right.
\end{equation}
which guarantees the antenna spacing and range constraints in~(\ref{eq:x-con-1}) and~(\ref{eq:x-con-2}).

\begin{figure}[t] 
   \centering
   \includegraphics[width=0.4\textwidth]{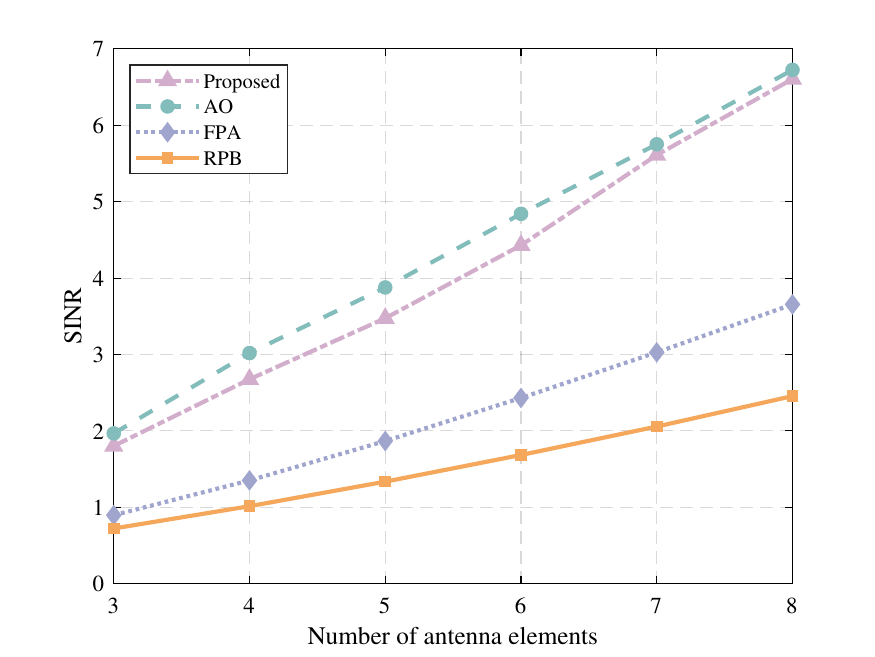} 
   \caption{Achieved SINR versus the number of antenna elements.}
   \label{fig_N}
\end{figure}

Based on the discussions on the optimizer design for receive beamforming and antenna element positioning, the overall learning architecture is shown in Fig.~\ref{fig:arch} and elaborated as follows. For the considered anti-jamming communication system, we first extract the directions of arrival from both legitimate source and jammers as $ \left\{ \theta_k \right\}_{k\in\mathcal{K}\cup\{0\}} $, which are then fed into the learning system. Through a fully-connected feature embedding module, the latent high-dimensional feature is exploited as the input of an MLP network to produce the set of ratios between minimum antenna spacing and the actual ones as $ \left\{ r_n \right\}_{n\in\mathcal{N}\backslash\{N\}} $. In consistence with the physical constraint on antenna spacing, we adopt $ \mathsf{ReLU} $ as the activation function at the input and hidden layers and $ \mathsf{Sigmoid} $ as the activation function of the output layer. Note the range of neural network output coincides with requirement of the antenna spacing ratio as $ r_n \in (0, 1] $, $ \forall n \in\mathcal{N}\backslash\{N\} $. The intermediate outputs are then converted into actual antenna positions by incorporating physical constraints including the minimum spacing between elements and the maximum allowable array length. The obtained antenna element positions are substituted into~(\ref{eq:AB}) to calculate the matrices $ \bm{A} $ and $ \bm{B} $, which are further used in solving the generalized Rayleigh quotient problem in~(\ref{eq:opt-w}) to derive the receive beamforming. Finally, the obtained receive beamforming, along with the antenna element positions, are fed into the loss function to conduct neural network training.

\begin{figure}[t]
   \centering
   \includegraphics[width=0.4\textwidth]{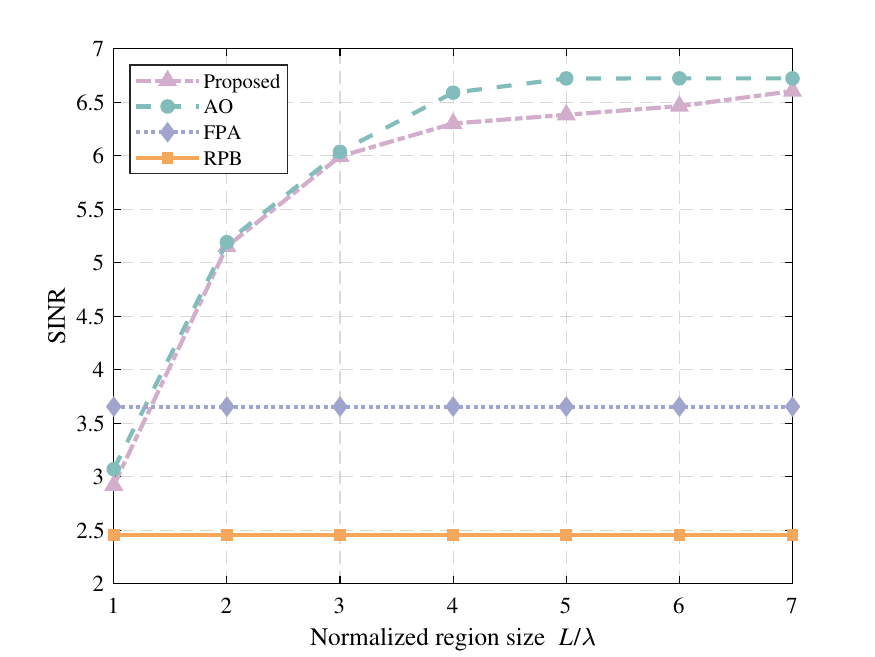} 
   \caption{Achieved SINR versus normalized region size.}
   \label{fig_L}
\end{figure}

\subsection{Loss Function and Training}

The proposed learning architecture intends to achieve effective anti-jamming transmissions. Denote the collective parameters of the neural network as $ \bm{\Theta} $, the learning system then adopts the input angles $ \left\{ \theta_k \right\}_{k\in\mathcal{K}\cup\{0\}} $, to output an anti-jamming strategy, denoted as $ \Phi\left( \left\{ \theta_k \right\}_{k\in\mathcal{K}\cup\{0\}}; \bm{\Theta} \right) $. The output strategy, including receive beamforming and antenna element positioning, targets to achieve the highest SINR as $ \eta\left( \Phi\left( \left\{ \theta_k \right\}_{k\in\mathcal{K}\cup\{0\}}; \bm{\Theta} \right); \left\{ \theta_k \right\}_{k\in\mathcal{K}\cup\{0\}} \right) $. Accordingly, we define the loss function of the neural network as the reciprocal of the SINR as
\begin{equation}
   \mathsf{L} \left(\bm{\Theta}\right) = \eta^{-1}\left( \Phi\left( \left\{ \theta_k \right\}_{k\in\mathcal{K}\cup\{0\}}; \bm{\Theta} \right); \left\{ \theta_k \right\}_{k\in\mathcal{K}\cup\{0\}} \right),
\end{equation}
for which the loss-minimized training approximates the highest transmission SINR. Moreover, we employ the unsupervised learning technique and the stochastic gradient descent is conducted for parameter updates.

As a further note, the proposed learning framework can be conveniently extended to the scenarios with multiple users. In this regard, we first need to update the input representation to handle the features of multiple legitimate users and jammers. Then, we can adopt the similar decomposition of receive beamforming and antenna position, where both parties can be approximated with a neural network module. Finally, the loss function can be defined to in consistence with the interested metric under multi-user scenarios. Therefore, the proposed approach manifests significant flexibility for movable antenna-based jamming mitigation in various scenarios.

Finally, the complexity of proposed approach is analyzed as follows. Generally, the learning framework consists of the modules for antenna positioning and beamforming. For antenna positioning, the complexity depends on the size of neural network. Assume $ L^\text{(e)} $ and $L^\text{(p)}$ hidden layers in the embedding and MLP modules, respectively, in addition to their individual input and output layers, with $ n^\text{(e)}_{i} $ and $ n^\text{(p)}_{i} $ specifying the number of neurons in the $i$-th layer, the complexity is then specified as $ \mathcal{O}\left(\sum\limits_{i=0}^{L^\text{(e)}}n^\text{(e)}_{i} n^\text{(e)}_{i+1} + \sum\limits_{i=0}^{L^\text{(p)}}n^\text{(p)}_{i} n^\text{(p)}_{i+1}\right) $. For the beamforming, the Rayleigh quotient is of a complexity of $ \mathcal{O}(N^3) $. Therefore, the overall training complexity is $ \mathcal{O}\left(L_{\text{tr}} L_{\text{sp}} \left(\sum\limits_{i=0}^{L^\text{(e)}}n^\text{(e)}_{i} n^\text{(e)}_{i+1} + \sum\limits_{i=0}^{L^\text{(p)}}n^\text{(p)}_{i} n^\text{(p)}_{i+1} + N^3\right)\right) $, when the neural network undergoes $ L_{\text{tr}} $ training epochs, each with $ L_{\text{sp}} $ training samples. Furthermore, when the trained network is used for inference involving only single forward procedures, the complexity is $ \mathcal{O} \left(\sum\limits_{i=0}^{L^\text{(e)}}n^\text{(e)}_{i} n^\text{(e)}_{i+1} + \sum\limits_{i=0}^{L^\text{(p)}}n^\text{(p)}_{i} n^\text{(p)}_{i+1} + N^3\right) $. Thus, the proposed learning framework overcomes the limitation of conventional approaches to tackle the complex, high-dimensional, and non-convex problem. The trained neural network can be efficiently exploited for fast inference with minimum computation overhead, allowing real-time adaptation to the adversarial environment.

\begin{figure}[t] 
   \centering
   \includegraphics[width=0.4\textwidth]{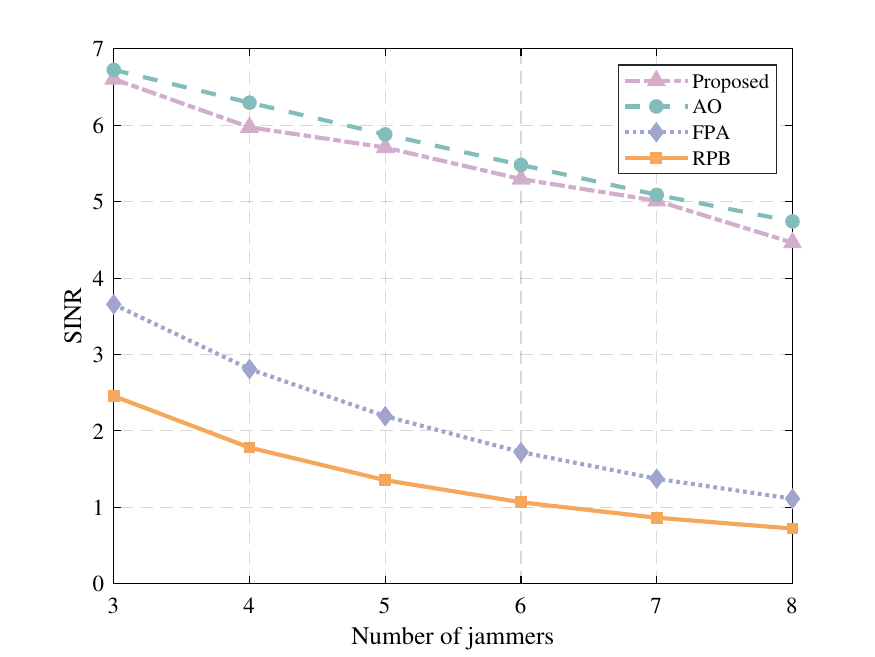} 
   \caption{Achieved SINR versus the number of jammers.}
   \label{fig_M}
\end{figure}

\section{Simulation Results}

In this section, we present the simulation results to validate the effectiveness of the proposed approach. The following simulation setup is used unless otherwise stated. We consider a network where the legitimate source is located with a randomly predefined direction of arrival, and the angle of jammers are randomly selected in accordance with a uniform distribution in $\left[0,\pi\right]$. There are 3 jammers, and the base station is with 8 movable antenna elements. The antenna elements can move within a range of 7 times of the wavelength, and the minimum spacing between adjacent elements is half wavelength. For the neural networks, there incorporate three fully connected layers. The model is trained over a dataset with 10\textsuperscript{5} samples, with a learning rate of 10\textsuperscript{-3} and a batch size of 10\textsuperscript{2}.

For performance comparison, the following schemes are used as benchmarks:
\begin{itemize}
   \item Alternating optimization (\textbf{AO}): the antenna element position and receive beamforming are obtained in an alternating optimization framework, following the similar process as in~\cite{sec-1}.
   \item Fixed-position antenna (\textbf{FPV}): conventional linear array with elements separated with two times of minimum spacing, receive beamforming obtained through Rayleigh quotient problem solving.
   \item Random positioning and beamforming (\textbf{RPB}): randomly determining the anti-jamming strategy while satisfying the constraints in~(\ref{eq:opt}).
\end{itemize}

\begin{figure}[t]
   \centering
   \includegraphics[width=0.4\textwidth]{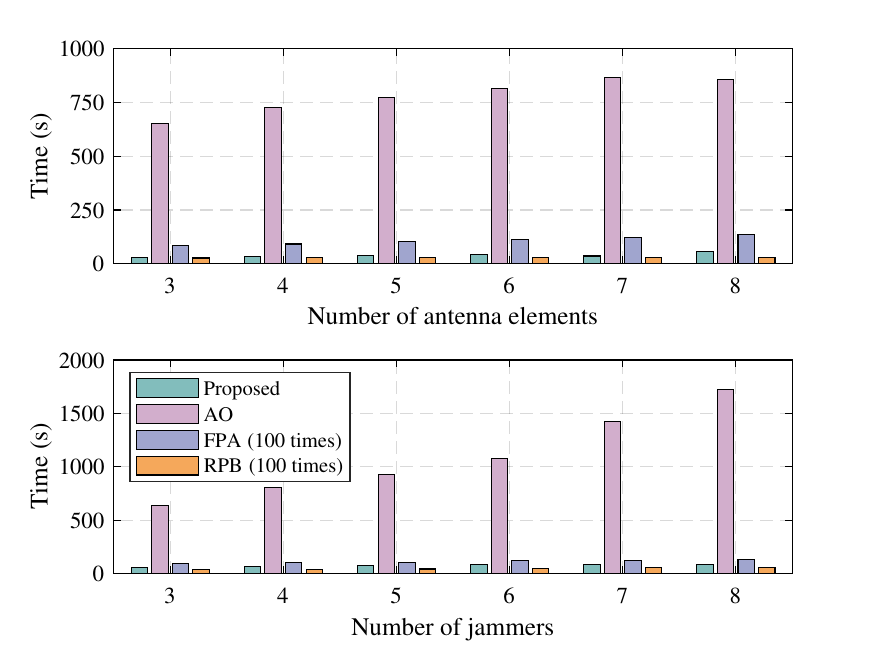} 
   \caption{Algorithm runtime.}
   \label{fig_T}
\end{figure}

In Figs.~\ref{fig_N},~\ref{fig_L}, and~\ref{fig_M}, we show the anti-jamming communication performance with different transmission scenarios. Specifically, in Fig.~\ref{fig_N}, we show the jamming mitigation performance with different numbers of movable antenna elements. Evidently, more antenna elements enable higher diversity and thus the achieved SINR is increased under all schemes. Compared with the optimization-based scheme, the proposed learning approach achieves comparable results. Moreover, as the flexible movement of antennas allow additional degree of freedom to combat jamming attacks, the anti-jamming performance gain with movable antennas as compared with fixed-position antennas becomes more evident with an increasing number of antennas. In Fig.~\ref{fig_L}, we evaluate the performance as the allowed region of antenna changes, where the region size is normalized with respect to the wavelength. Similarly, the performance of our proposal approaches that under optimization. Also, as the allowed region becomes larger, the anti-jamming performance can be further enhanced, yet the achieve SINR also tends to be plateaued when the region size becomes sufficiently large. In contrast to the case with fixed-position antenna, we can see that the movable antenna can more effectively alleviate the jamming attacks even within smaller allowed regions. In Fig.~\ref{fig_M}, we demonstrate the jamming mitigation with respect to the number of jammers. As expected, when there are more jammers and the attacks become more severe, the transmission performance in terms of achieved SINR becomes degraded. Also, the performance of proposed learning scheme approaches that obtained through optimization, regardless of the number of jammers. Moreover, the flexible movement leads to a much higher transmission SINR as compared with the scenarios with fixed-position antenna. While the proposed deep learning framework demonstrates near-optimal performance, some gaps can be spotted compared to the optimization-based method. We may potentially resort to further fine-tuning, dataset enrichment, etc., to enhance the performance.

In Fig.~\ref{fig_T}, we evaluate the runtime efficiency of different schemes. For our proposal, when the established learning network is well trained, it is capable for prompt calculation and execution to provide the anti-jamming policy. Accordingly, the time consumption remains low regardless of the number of antenna elements and jammers. For the case with optimization-based approach, much more time is required to reach some suboptima with an acceptable precision. Particularly, when there are more jammers and thus the communication environment becomes more complicated, the time consumption by adopting optimization-based method increases evidently. For the cases under FPV and RPB, they are of nearly closed-form calculations and the time taken to run 100 times is depicted for reference.

\section{Conclusion}

In this paper, we have investigated the movable antenna-assisted jamming mitigation. We have adopted a deep learning-assisted approach to achieve anti-jamming strategy design by jointly considering the receive beamforming and antenna element positioning. The results have indicated that our proposal achieves near-optimal performance while producing the anti-jamming policy in a significantly more prompt manner. Therefore, this paper validates the potential of movable antennas in jamming mitigation and the proposed approach is capable for efficient anti-jamming strategy determination in various wireless environments.

\bibliographystyle{IEEEtran}
\bibliography{main}

\end{document}